\documentclass[aps, pre, twocolumn,showkeys,showpacs,superscriptaddress]{revtex4-1}

\usepackage[nottoc,numbib]{tocbibind}
\usepackage[T1]{fontenc}
\usepackage[utf8]{inputenc}
\usepackage{amssymb}
\usepackage{pifont}
\usepackage{textcomp}
\usepackage[centertags]{amsmath}
\usepackage{graphicx}
\graphicspath{{figures/}{logos/}}
\DeclareGraphicsExtensions{.pdf,.png,.jpg}
\usepackage{booktabs}
\usepackage{url}
\usepackage{epsfig}
\usepackage{textcomp}
\usepackage{subfigure}
\usepackage{lmodern}
\usepackage[squaren]{SIunits}
\usepackage{placeins}
\usepackage{color}
\usepackage{listings}
\usepackage{floatpag}
\usepackage{needspace}
\usepackage[]{natbib}
\usepackage{verbatim}
\usepackage{array}

\newcommand{\bcdot}{\hspace{-3.5pt}\ensuremath{\cdot}\hspace{-3.5pt}}
\newcommand{\pder}[2]{\ensuremath\frac{\partial #1}{\partial #2}}

\newcommand{\Rot}[1]{\ensuremath{ \nabla \times #1 }}
\newcommand{\Div}[1]{\ensuremath{ \nabla \cdot #1 }}

\newcommand{\abs}[1]{\left|#1\right|}
\newcommand{\mkl}[1]{\left\{#1\right\}}
\newcommand{\kl}[1]{\ensuremath{\left(#1\right)}}
\newcommand{\ekl}[1]{\ensuremath{\left[#1\right]}}

\makeatletter
\newcommand{\thickhline}{%
    \noalign {\ifnum 0=`}\fi \hrule height 1pt
    \futurelet \reserved@a \@xhline
}
\newcolumntype{"}{@{\hskip\tabcolsep\vrule width 1pt\hskip\tabcolsep}}
\makeatother

\begin{document}


\title{Hall Effect in Diluted Plasmas}


\author{N. Offeddu} \email{nico.offeddu@gmail.com} \affiliation{ ETH
  Z\"urich, Computational Physics for Engineering Materials, Institute
  for Building Materials, Wolfgang-Pauli-Strasse 27, HIT, CH-8093 Z\"urich
  (Switzerland)}

\author{M. Mendoza} \email{mmendoza@ethz.ch} \affiliation{ ETH
  Z\"urich, Computational Physics for Engineering Materials, Institute
  for Building Materials, Wolfgang-Pauli-Strasse 27, HIT, CH-8093 Z\"urich
  (Switzerland)}



\date{\today}

\begin{abstract}
We study the Hall effect in diluted plasmas within the two-fluids theory. Composed by two distinct species with opposite charge, such as electrons and ions in fully ionised hydrogen, the plasma is driven by an electric field through a channel in the presence of a transversal magnetic field. As a consequence, a separation of charge is induced producing an electric potential difference. We have found a general relation for the Hall voltage as function of the mass and viscosity ratios, which converges to the usual expression in the limit of solid matter, i.e. when ions are much more massive than electrons. All the simulations have been performed using a three-dimensional Lattice-Boltzmann model, which has been also validated for some relevant applications. Finally, we discuss the importance of our findings in the light of recent developments in plasma physics, in particular in magnetic reconnection.
\end{abstract}


\maketitle  

\section{Introduction}
It is a well known and documented phenomena that plasma flow is affected by magnetic fields. Extensive research has been done on the Hall effect \cite{hall,hall_paper}. However, most of these studies, e.g. \cite{hall_in_plasma_example}, focus on a plasma flow driven by pressure gradients, i.e., all the particles are accelerated in the same direction. Transversal magnetic fields therefore tend to separate charges and to create a voltage across the streamlines. However, if the flow is driven by an electric field, the particles are accelerated according to their charge and the plasma flow can take non-trivial configurations.

To understand the expected behaviour of a plasma flowing in the presence of a magnetic field we briefly recall the assumptions made in the derivation of the classic Hall effect in conductors \cite{hall_classic}: the electrical conductivity $\sigma_0$ is a constant scalar in a homogeneous and isotropic material; and the current is purely composed by electrons confined in a conductor. In Ohmic conductors, the Hall effect can be reduced to the Lorentz force acting on moving charge carriers, balanced by the electric potential which is eventually generated by their drift. In a fluid, the macroscopic effects can be ascribed to the interactions of the electromagnetic fields with the particles of which the fluid is composed of. While considering plasmas, the assumptions made in the derivation of the classic Hall effect change slightly. First of all, the conductivity is not a constant given by the material's properties and an alternative description has to be used. Regarding the second assumption, even if ions have a mass of three orders of magnitudes larger than the one of the electrons, they also carry charge, move, and can therefore introduce corrections. Besides this, the charge carriers in plasma are not always only protons and electrons, e.g., the plasma might not be fully ionised or be composed by different species of ions. A generalised approach to the Hall effect is needed. 

To our knowledge the hall effect in two-fluid plasma flows driven by electric fields has not been investigated to date. Due to its importance, particularly in astrophysics, and its simple composition we will consider a fully ionised plasma. Our analysis is done on a fluid with two  populations of opposite charge and different mass, as in the case of ionised hydrogen, without restricting the model to this specific case. The flow is driven by an electric field, thus the populations flow in opposite directions and the drift velocities are affected in a non trivial way. Both the velocities of electrons and ions will be reduced by Coulomb collisions, and only a weaker voltage can be built up across the flow. A second non trivial effect is that the moving protons will also feel the Lorentz force. Since the sign of their velocity and their charge are opposite to those of the electrons, they will be accelerated in the same direction as the electrons, further reducing the voltage. As a special example, if negative and positive charge carriers (having charge $q_-$ and $q_+$, respectively) have the same mass and viscosity, the magnitude of their drift velocity ($\vec{v}_-$ and $\vec{v}_+$, respectively) due to the external field $\vec{B}$  will be the same, as well as the Lorentz force, $F_L^+ = q_+ \vec{v}_+ \times \vec{B} = (-q_-) (-\vec{v}_-) \times \vec{B} = F_L^-$. In such a case, the two species are equally accelerated in the same direction, with the result that the Hall effect (in the classical sense of a current-induced voltage) is completely suppressed. The flow will still be affected, because of the pressure gradient in the direction in which the two populations are accelerated. This has important consequences, especially concerning  the stratification of populations and pressure gradients, which might affect the flow in systems with high Reynolds number \cite{flow_separation}.

To perform our study, we will use the lattice Boltzmann model reported in Ref.~\cite{miller_bibbia}. This model is based on the two-fluids theory for plasmas and has been successfully used to study magnetic reconnection processes. Part of the aim of our work is to extend the range of possible applications of the model, including some new validation tests, e.g. Taylor-Green vortices, propagation of electromagnetic waves, and the standard Hall effect in conductors.

The paper is structured as follows: in  Sec. \ref{sec:two_fluid} the theoretical model used in this work is presented. In Sec. \ref{subsec:LBmodel} the numerical model used for the simulations is outlined and is then validated in Sec. \ref{sec:validation} through comparisons with systems with a known analytical solution. Finally, in Sec. \ref{subsec:Hall_in_plasma}, our results are analysed and discussed.

\section{Two-fluid theory for plasmas}
\label{sec:two_fluid}
Ideal and resistive magnetohydrodynamics (MHD) rely on the assumption that the electric field within the plasma is negligible \cite{plasma_general}. If the conductivity is high enough and the time scales of the processes are long enough, the charges in the plasma can easily rearrange and cancel the internal electric field out. However, if the temperature of the plasma is high enough or the plasma is highly diluted, the conductivity drops (as we will see in Eq. \ref{eq:sigma_0}) and electron and ion momenta must be taken into account separately \cite{plasma_general}. A possible approach is considering the plasma as composed by two fluids. This approach converges towards MHD when the mentioned conditions are relaxed. In this section, the procedure is briefly outlined and the main two-fluid equations are introduced.

The conservation of momentum for each fluid component in the presence of electric ($\vec{E}$) and magnetic ($\vec{B}$) fields is governed by the following equation \cite{plasma_two_fluid},
\begin{align}
\label{eq:gen_mom_eq}
m_s n_s \kl{\pder{\vec{v}_s}{t} + \kl{\vec{v}_s\bcdot\nabla}\vec{v}_s} &=  
n_s q_s\kl{\vec{E} + \vec{v}_s\times\vec{B}} \nonumber \\
&- \nabla P_s + n_s\eta_s \nabla^2 \vec{v}_s \nonumber \\
& -\nu\rho_0\kl{\vec{v}_s - \vec{v}_{\bar{s}}} \quad ,
\end{align}
where $n_s$, $m_s$, $\vec{v}_s$, $q_s$, $P_s$, and $\eta_s$ are the particle density, mass, fluid velocity, electric charge, pressure, and kinematic viscosity, respectively, for each of the particle species $s=0,1$. The loss of momentum density of one population due to collisions with the other one is described by the last term at the right hand side of Eq.~\eqref{eq:gen_mom_eq}, where $\nu$ is the collision frequency and $\bar s \equiv  \kl{s+1}\text{mod}(2)$. Since the momentum transfer is symmetric, the same relation is valid for both populations. External forces can be considered by adding an extra term. Note that we have assumed non-relativistic regimes, i.e.,  where the characteristic fluid velocities are much smaller than the speed of light. Under this assumption, we have used the non-relativistic fluid equations. 

Combining the continuity equation and the momentum equation of the two populations, and defining the new macroscopic variables
\begin{align}
  \begin{aligned}
    \text{mass density}\hspace{0.5cm}
    \rho_m &= \sum_s{n_sm_s} \quad , \\
    \text{charge density}\hspace{0.5cm}
    \rho_q &= \sum_s{q_sn_s} \quad , \\
    \text{total velocity}\hspace{0.5cm}
    \vec{v} &= \frac{1}{\rho_m}\sum_s{n_sm_s\vec{v}_s} \quad ,\\
    \text{current density}\hspace{0.5cm}
    \vec{j} &= \sum_s{n_sq_s\vec{v}_s} \quad ,\\
    \text{pressure}\hspace{0.5cm}
    P &= \sum_s{P_s},\\
  \end{aligned}
  \label{eq:new_var}
\end{align}
we obtain the total mass and charge conservation equations
\begin{align}
  \begin{aligned}
    0 &=\pder{\rho_m}{t} + \nabla\bcdot \kl{\rho_m\vec{v}} \quad , \\
    0 &=\pder{\rho_q}{t} + \nabla \bcdot\vec{j} \quad . \\
  \end{aligned}
  \label{eq:new_conserv}
\end{align}
If polytropic processes are assumed, an equation of state of the form
\begin{equation}
\sum_i P_sn_s^{-\gamma_s} = \text{constant}
\label{eq:eq_of_state1}
\end{equation}
is used. This restricts the model to processes in which the energy transfer ratio of heat to work  $\frac{\delta Q}{\delta W}$ is constant. The value of $\gamma$ is dependent on the physical system. For the three-dimensional isothermal limit we have that $\gamma = 1$. In the limit of quasi-neutrality ($n_0 \approx n_1$) we have the the same $\gamma$ for all the populations and  the state equation simplifies to
\begin{equation}
P_s n_s^{-\gamma_s} = \text{constant}.
\label{eq:eq_of_state2}
\end{equation}
Combining the momentum and continuity equations for the two populations, the generalized Ohm's law can be derived \cite{plasma_general_2} as:
\begin{align}
\vec{E} + \vec{v}\times\vec{B} 
&= - \frac{m_0 m_1}{q_0 q_1} \pder{\,}{t} \kl{\frac{\vec{j}}{\rho_m}} &\kl{\text{\emph{electron inertia}}  }\nonumber \\
& + \frac{1}{\rho_m} \kl{\frac{m_1}{q_1}+\frac{m_0}{q_0}}\vec{j}\times\vec{B}   &\kl{\text{\emph{Hall term}}  } \nonumber \\
& - \frac{m_0 m_1}{\rho_m q_0 q_1}\kl{\frac{q_1}{m_1}\nabla P_1 + \frac{q_0}{m_0}\nabla P_0 }  \hspace{-0.7cm} &\kl{P_1\text{\emph{ gradient}}  } \nonumber \\
& - \kl{1 - \frac{q_0 m_0}{ q_1 m_1}} \frac{m_0 m_1}{\rho_m q_0 q_1} \nu \, \vec{j} \quad ,   &\kl{\text{\emph{Ohm's term}}  } \nonumber \\
\label{eq:general_ohm}
\end{align}
where we have neglected the viscous terms in Eq.~\eqref{eq:gen_mom_eq}. In most cases, one does not deal with the full generalized equation, but rather employes simplified approximations. For example, since $\frac{q_1}{m_1}<<\frac{q_0}{m_0}$, the term with the ionic pressure gradient is much smaller than the one with the electronic pressure gradient, and it is often neglected. Note that in a steady state, in the limit of quasi-neutrality and in absence of magnetic fields perpendicular to the current, the generalized Ohm's law simplifies to
\begin{equation}
\label{eq:simpl_ohm}
\vec{E} =  -\frac{m_0 m_1}{q_0 q_1 \rho_m}  \nu \vec{j} 
        \hspace{0.2cm}\overset{m_1 >> m_0}{\approx}\hspace{0.2cm}  \underbrace{ \frac{m_0 \nu}{q_0^2 n_0 }  }_{\text{resistivity}} \vec{j}
        = \frac{1}{\sigma_0}\,\,\vec{j} \quad .
\end{equation}
This means that in an ionized plasma, if the collision frequency of the electrons and ions is much higher than the typical gyro frequencies, the conductivity can be regarded as a scalar given by \cite{fridman}
\begin{equation}
\label{eq:sigma_0}
\sigma_0 = \frac{n_0 q_0^2}{m_0\nu} \quad ,
\end{equation} 
where $n_0$ is the electron density, $q_0$ the electric charge, $m_0$ the electron mass and $\nu$ the frequency of collision between the charge carrier species. Intuitively, if the electrons and ions collide more often, the conductivity will be suppressed. 

Finally, the fluid equations contain interactions with the electromagnetic fields, which are governed by the Maxwell equations:
\begin{align}
  \begin{aligned}
    \Rot{\vec{E}} &=-\pder{\vec{B}}{t} \quad ,&
    \Div{\vec{B}} &=0 \quad ,
    \\
    \Rot{\vec{B}} &=\mu_0\vec{j} + \mu_0\epsilon_0\pder{\vec{E}}{t} \quad ,&
    \Div{\vec{E}} &=\frac{1}{\epsilon_0}\rho_{q} \quad ,
  \end{aligned}
  \label{eq:maxwell}
\end{align}
where $\epsilon_0$ and $\mu_0$ are the electric permittivity and magnetic permeability of vacuum. 

The generalized Ohm's law, Eq. \eqref{eq:general_ohm}, the momentum equation, Eq. \eqref{eq:gen_mom_eq}, the equation of state, Eq. \eqref{eq:eq_of_state1}, and the Maxwell equations, Eq. \eqref{eq:maxwell}, complete the set of equations within the two-fluid theory, which are also recovered by the numerical solver in Ref.~\cite{miller_bibbia}.

\section{\label{subsec:LBmodel}Lattice Boltzmann Model}

The lattice Boltzmann method \cite{succi} has been already successfully applied in the study of plasma physics and magnetohydrodynamics \cite{lb_mhd_vecchio,two_dim_mhd}. In this work it is applied to the two-fluid theory, modeling the conductivity via a collision parameter $\nu$. The negative charge carriers are labelled by $s=0$ and the positive ones by $s=1$. Following Ref. \cite{miller_bibbia,miller_2}, a D3Q19 ($3$ dimensions and $19$ vectors) lattice has been used for all the simulations. Space is divided into a regular 3D grid and all the quantities used in this work are given in lattice units. At each lattice site, distributed on three perpendicular planes, there are 19 independent velocity vectors  $\hat{v}_i$ weighted by factors $\omega_i$. Together with the rest vector ($\hat{v}_0 = \mkl{0,0,0}$, weighted by $\omega_0=\frac{1}{3}$), these are given by 12 vectors of magnitude $\sqrt{2}$, 
\begin{align}
\hat{v}_i &= \sqrt{2} \mkl{\cos\ekl{\kl{2i-1}\frac{\pi}{4}},\sin\ekl{\kl{2i-1}\frac{\pi}{4}},0} \quad , \nonumber \\
\hat{v}_{i+4} &= \sqrt{2} \mkl{\cos\ekl{\kl{2i-1}\frac{\pi}{4}},0,\sin\ekl{\kl{2i-1}\frac{\pi}{4}}} \quad , \nonumber \\
\hat{v}_{i+8} &= \sqrt{2} \mkl{0,\cos\ekl{\kl{2i-1}\frac{\pi}{4}},\sin\ekl{\kl{2i-1}\frac{\pi}{4}}} \quad , \nonumber \\
\end{align}
for $i\in\mkl{1,2,3,4}$, weighted by $\omega_{1-12}=\frac{1}{36}$  and 6 more vectors of magnitude 1 
\begin{align}
\hat{v}_{12+i} &= \mkl{\kl{-1}^{i},0,0} \quad , \nonumber \\
\hat{v}_{14+i} &= \mkl{0,\kl{-1}^{i},0} \quad , \nonumber \\
\hat{v}_{16+i} &= \mkl{0,0,\kl{-1}^{i}} \quad , \nonumber \\
\end{align}
for $i\in\mkl{1,2}$, with $\omega_{13-18}=\frac{1}{18}$  (See Fig. \ref{fig:D3Q19}). Associated to each of these velocity vectors there are two distribution functions $f_i^s$ for the populations.

There are also $25$ distributions $G_i$ for the electric and magnetic fields associated to field vectors $\hat{e}_i^s$ and $\hat{b}_i^s$. For $i\in\mkl{1,2,3,4}$, the electric vectors on the three planes are defined by
\begin{align}
\hat{e}_{i}^0 &= \frac{1}{2}\hat{v}_{\kl{i+2}\text{mod}4+1} \quad , \nonumber \\
\hat{e}_{i+4}^0 &= \frac{1}{2}\hat{v}_{\kl{i+2}\text{mod}4+5} \quad , \nonumber \\
\hat{e}_{i+8}^0 &= \frac{1}{2}\hat{v}_{\kl{i+2}\text{mod}4+9} \quad , \nonumber \\
\end{align}
with $\hat{e}_{0}^0=\hat{v}_0$ and $\hat{e}_{i}^1 = -\hat{e}_{i}^0$. The magnetic vectors are perpendicular to the associated velocity and electric vectors:
\begin{equation}
\hat{b}_{i}^s = \hat{v}_{i}\times\hat{e}_{i}^s.
\end{equation}
Of the 25 electric vectors, only 13 are independent. Of the 25 magnetic vectors, only 7 are independent.

The macroscopic variables for the fluid (particle density, momentum density, electric and magnetic fields) at each cell can be computed as follows:
\begin{figure}
	\center
    \includegraphics[width=0.5\textwidth]{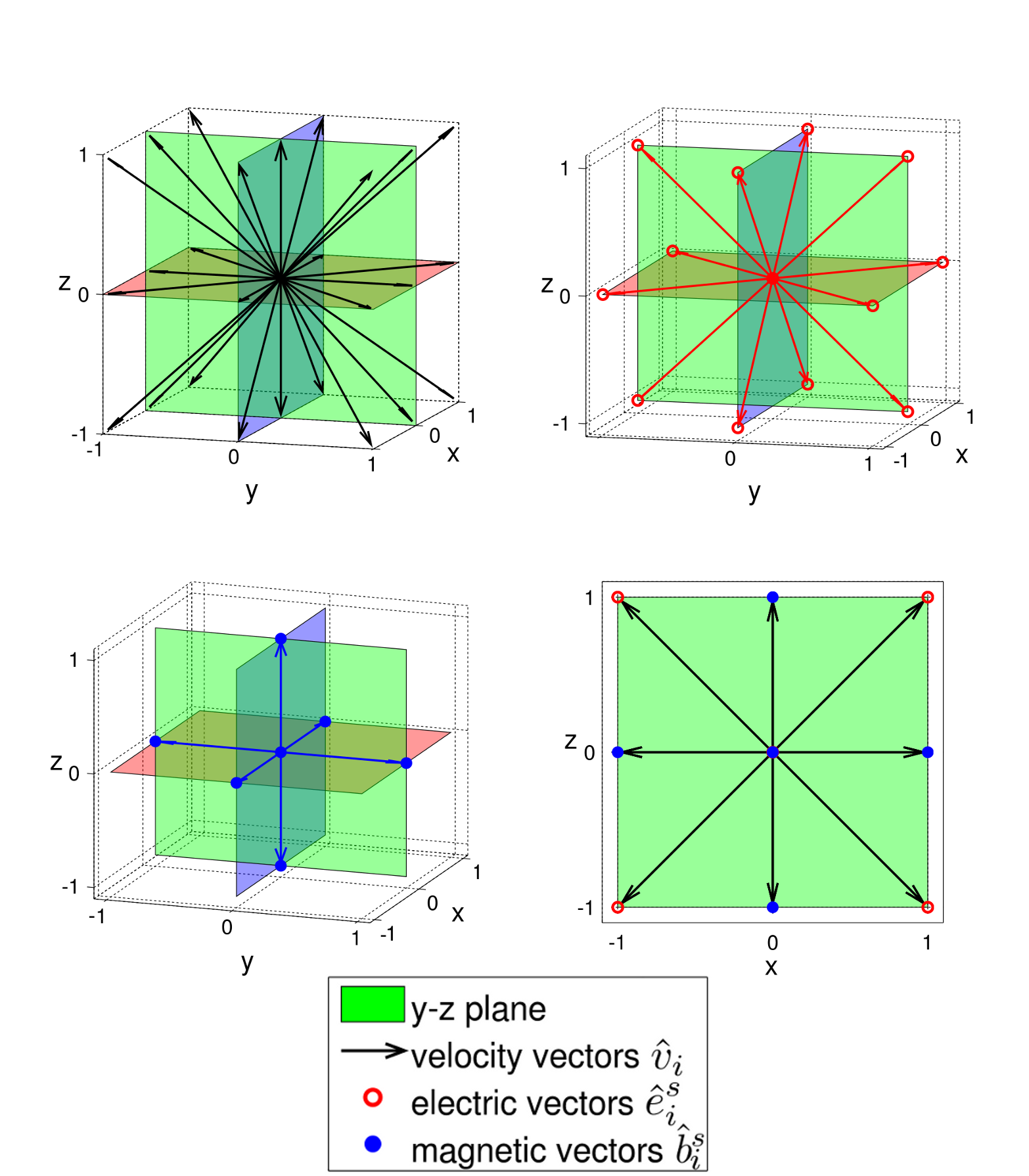}
  \caption{\textbf{Lattice velocities scheme.} Lattice velocities $\hat{v}_i$, to which the distributions $f_i^s$ are associated. There are 9 velocity vectors per plane (including the rest vector), for a total of 19 independent vectors.}
\label{fig:D3Q19}
\end{figure}
\begin{align}
\rho_s &= \sum_i f_i^s \quad ,           &  \rho_s \vec{v}_s &= \sum_i f_i^s \hat{v}_i \quad ,    \nonumber \\
 \vec{E} &= \sum_{i,s} G_i^s \hat{e}_i^s \quad ,       &  \vec{B} &= \sum_{i,s} G_i^s \hat{b}_i^s \quad .   
\end{align}
where $\rho_s = n_s m_s$. From these, the total charge density and the current can be calculated as:
\begin{align}
\rho_c &=  \sum_s \frac{q_s}{m_s}\rho_s \quad ,           
& \vec{j} &=  \sum_s \frac{q_s}{m_s}\rho_s\vec{v}_s \quad .     
\end{align}
For every charge carrier species, every volume cell is under the influence of the Lorentz force, external forces and the collision with the other species. According to Eq. \eqref{eq:gen_mom_eq}, the total force on a volume element can be written as 
\begin{equation}
 \vec{F}_s = \frac{q_s}{m_s} \rho_s \kl{\vec{E} + \vec{v}_{s}\times \vec{B}}
				- \nu \rho_0 \kl{\vec{v}_{s} - \vec{v}_{\bar s} } + \vec{F}_s^{\text{ext.}},
\label{eq:phys_force}
\end{equation}
with $ \vec{F}_s^{\,\text{ext.}}$ an external force. For the forcing, the model proposed by \citet{guo_forcing} has been adopted. The populations $f^s_i$ and $G_i$ are updated according to
\begin{align}
f^s_i\ekl{\vec{x} + \hat{v}_i,t+1} &= \kl{1-\frac{1}{\tau_s}} f^s_i\ekl{\vec{x},t} + \frac{1}{\tau_s}   f^{s,\text{eq}}_i\ekl{\vec{x},t}  + T^s_i,         \nonumber \\
G_i\ekl{\vec{x} + \hat{v}_i,t+1} &= \kl{1-\frac{1}{\tau_G}} G_i\ekl{\vec{x},t} + \frac{1}{\tau_G}   G^{\text{eq}}_i\ekl{\vec{x},t}  + T^G_i,         
\end{align}
where $\tau_s$ and $\tau_G$ are the relaxation times for the two fluids and the electromagnetic distribution functions. $ T^s_i$ and  $ T^G_i$ are forcing terms to be determined with
\begin{align}
\hspace{-0.2cm}T^s_i &= \kl{1-\frac{1}{2\tau_s}} \omega_i\kl{ 3\kl{\hat{v}_i-\vec{v}_s\,'}\bcdot \vec{Z}_s  + 9\kl{\hat{v}_i\bcdot\vec{v}_s\,'}(\hat{v}_i\bcdot\vec{Z}_s ) },      \nonumber \\
T^G_i &= 0. \nonumber \\     
\end{align}
The corrected velocities $\vec{v}_{s}\,'$ and the forcing term $\vec{Z}_s$ can be calculated up to a precision of second order in the expansion of velocities \cite{miller_bibbia} starting by the physical force given in Eq. \eqref{eq:phys_force}:
\begin{equation}
	\vec{Z}_s = \frac
				{\kl{1 + \frac{q_{\bar s}^2}{8 m_{\bar s}^2}\rho_{\bar s}\mu_0}\vec{F}_s}
				{1 + \frac{1}{8}\mu_0 \kl{
									\frac{q_{ s}^2}{ m_{ s}^2} \rho_{ s}
								+	\frac{q_{\bar s}^2}{ m_{\bar s}^2} \rho_{\bar s}
											}
				}
			- 	\frac
			{\kl{\frac{q_{\bar s}q_{s}}{8 m_{\bar s}m_{s}}\rho_{s}\mu_0}\vec{F}_{\bar s}}
				{1 + \frac{1}{8}\mu_0 \kl{
									\frac{q_{ s}^2}{ m_{ s}^2} \rho_{ s}
								+	\frac{q_{\bar s}^2}{ m_{\bar s}^2} \rho_{\bar s}
											}
				}.
\end{equation}
With these expressions, the corrected velocities, electric field and currents can be calculated as
\begin{equation}
	\vec{v}_s{\,}' = \vec{v}_s + \frac{\vec{Z}_{s}}{2\rho_s} \quad ,
\end{equation}
\begin{equation}
	\vec{E}\,' = \vec{E} - \frac{1}{4}\mu_0 \vec{j}\,' \quad ,
\label{eq:corr_e_field}
\end{equation}
with the corrected current
\begin{equation}
	\vec{j}\,' = \vec{j} + \sum_s {\frac{q_s}{m_s} \kl{\frac{1}{2}\vec{Z}_{s}}} \quad ,
\end{equation}
that plugged into Eq. \eqref{eq:corr_e_field} yields
\begin{equation}
	\vec{E}\,' = \vec{E} - \frac{1}{4}\mu_0 \kl{\vec{j} + \sum_s {\frac{q_s}{m_s} \kl{\frac{1}{2}\vec{Z}_{s}}}} \quad .
\end{equation}
The equilibrium distribution functions for the fluids populations can be calculated by using Hermite polynomials, $\mathcal{H}_l$:
\begin{equation}
f_i ^ {s,\text{eq}} \kl{\vec{r},t} =\omega_i \kl{
						\frac{1}{c_s^0}a_0^s\mathcal{H}_0
						+ \frac{1}{c_s^2}a_1^s\mathcal{H}_1
						+ \frac{1}{2 c_s^4}a_2^s\mathcal{H}_2
}.
\end{equation}
The expansion coefficients can be calculated by the projections
\begin{align}
a_0^s &= \int f_i ^ {s,\text{eq}} \mathcal{H}_0 = \rho_s  \quad ,\\
a_1^s &= \int f_i ^ {s,\text{eq}} \mathcal{H}_1 = \rho_s \vec{v}_s  \quad ,\\
a_2^s &= \int f_i ^ {s,\text{eq}} \mathcal{H}_2 = \int   f_i ^ {s,\text{eq}} v^\alpha v^\beta 
					-c_s^2\int f_i^{s,\text{eq}}\nonumber \\ 
 &= \kl{P_s-c_s^2 \rho_s}\delta^{\alpha,\beta} + \rho_s v^\alpha v^\beta  \quad .
\end{align}
And, with an analogous treatment of the electromagnetic fields, this yields the following equilibrium distributions:
\begin{subequations}
\begin{align}
	\label{eq:f_i_equilibria}
	\begin{split}
	f_i ^ {s,\text{eq}} \kl{\vec{r},t} &=\omega_i 
		\large{[} 
			\rho_s 
			+ \frac{1}{c_s^2}\rho_s\vec{v}_s\bcdot \hat{v}_i 
			+ \frac{1}{2 c_s^4}\kl{P_s-c_s^2\rho_s}\hat{v}_i^2 \\
			&\quad{} + \frac{1}{2 c_s^4}\rho_s\kl{\vec{v}_s\bcdot\hat{v}_i }^2 
			- \frac{3}{2 c_s^2}\kl{P_s-c_s^2\rho_s} \\
			&\quad{} - \frac{1}{2 c_s^2} \rho_s \vec{v}_s^2    
		\large{]} \quad , 
	\end{split}\\
	\label{eq:G_i_equilibria}
	G_i ^ {\,\text{eq}} \kl{\vec{r},t} 
					&= 		\frac{1}{4} \vec{E}\,'\bcdot  \hat{e}_i
						+	\frac{1}{8} \vec{B}\bcdot  \hat{b}_i \quad . 
\end{align}
\label{eq:equilibria}
\end{subequations}
This model recovers the incompressible and viscous fluid equation  
\begin{align}
\label{eq:two_fluid_navier}
\rho_s \kl{\pder{\vec{v}\,'_s}{t} + \kl{\vec{v}\,'_s\bcdot\nabla}\vec{v}\,'_s} = 
&\frac{q_s}{m_s}\rho_s\kl{\vec{E} + \vec{v}\,_s\times\vec{B}}\nonumber \\
&- \nabla P_s \nonumber \\
&+\rho_s\eta_s\nabla^2\vec{v}\,'_s \nonumber \\
&- \nu\rho_0\kl{\vec{v}\,'_s - \vec{v}\,'_{\bar s}} \nonumber \\
&+ \vec{F}\,^{\text{ext.}}_s, \nonumber \\
\end{align}
with the kinematic viscosity 
\begin{equation}
\eta_s = \kl{\tau_s - \frac{1}{2}} c_s^2 \delta t  \quad .
\label{eq:kin_visc}
\end{equation}
with a speed of sound $c_s=\frac{1}{\sqrt{3}}$ for $\frac{\delta x}{\delta t}=1$. All the values given in this work are in lattice units and can be converted real units via dimensionless quantities, e.g. Reynolds number.

\section{\label{sec:validation}Validations}

To validate the algorithm, various tests are presented in this section. Combinations of external and internal forces are compared to analytical results. Doubling the system size or changing the direction of flows or electromagnetic field propagation does not affect any of the results.

\subsection{Taylor-Green vortices}
\label{sec:taylor_green}
\begin{figure}
		\centering
		\includegraphics[width=0.4\textwidth]{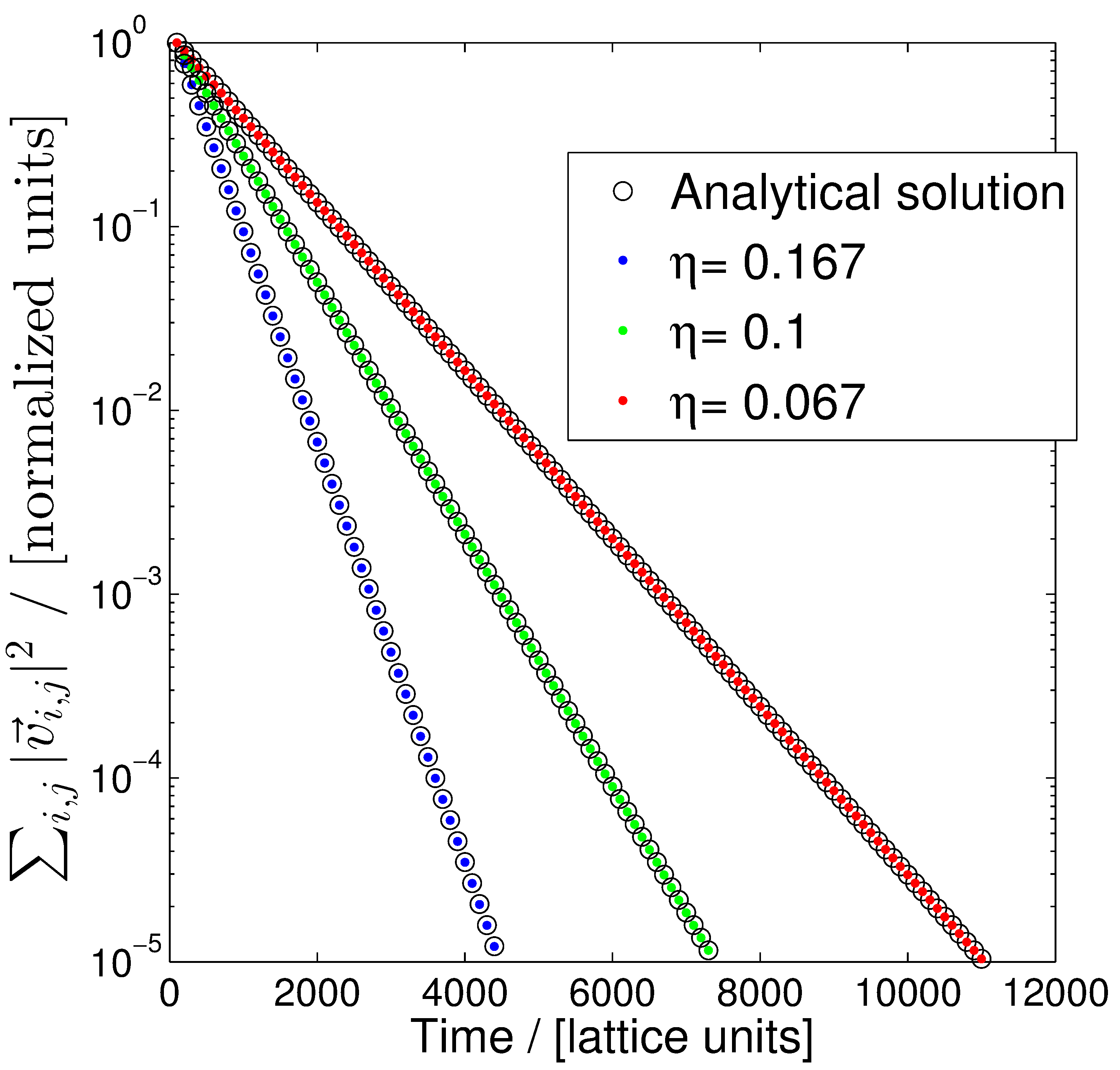}
		\caption{\textbf{Energy decay in Taylor-Green vortices.} Semi-log plot of the normalized energy over time of a $100\times100\times1$ grid with an initial configuration as defined in Eq. \eqref{eq:taylor_in_profile}. The slope in the semi-log plot is in agreement with the analytical results.}
	\label{fig:taylor}
\end{figure}

To test the effect of viscosity, the exponential decay in the velocity field of a Taylor-green vortex has been examined. As an initial configuration a velocity field of 
\begin{align}
\label{eq:taylor_in_profile}
v_x &= V_0\sin\ekl{k_x\,i}\cos\ekl{k_y\,j} \quad , \nonumber  \\
v_y &= -V_0\cos\ekl{k_x\,i}\sin\ekl{k_y\,j} \quad , \nonumber  \\
v_z &= 0 \quad ,
\end{align}
has been imposed on a grid of size $100\times100\times1$. Here, $V_0$ is a constant, $i$ and $j$ are chosen such that $0\le i,j< 100$ and $k_x=k_y = \frac{2\pi}{100}$. After \citet{taylor_green}, the total kinetic energy of the populations decays as $\text{exp}\ekl{-2\eta_s\kl{k_x^2 + k_y^2} t}$, assuming $\nu=0$, with $\eta_s$ being the kinematic viscosity of the respective population. The semi-log plot of the normalized sum of all the squared velocities over all sites, i.e.
\begin{equation}
\sum_{\text{sites}}|\vec{v}|^2 \quad ,
\end{equation}
can be seen in Fig. \ref{fig:taylor}. The effect of the viscosity on the exponent can be studied by varying $\tau_s$, see Eq.\eqref{eq:kin_visc}. 

\subsection{Propagation of EM waves}
 \label{sec:propagation_EM}

\begin{figure}
		\centering
		\includegraphics[width=0.5\textwidth]{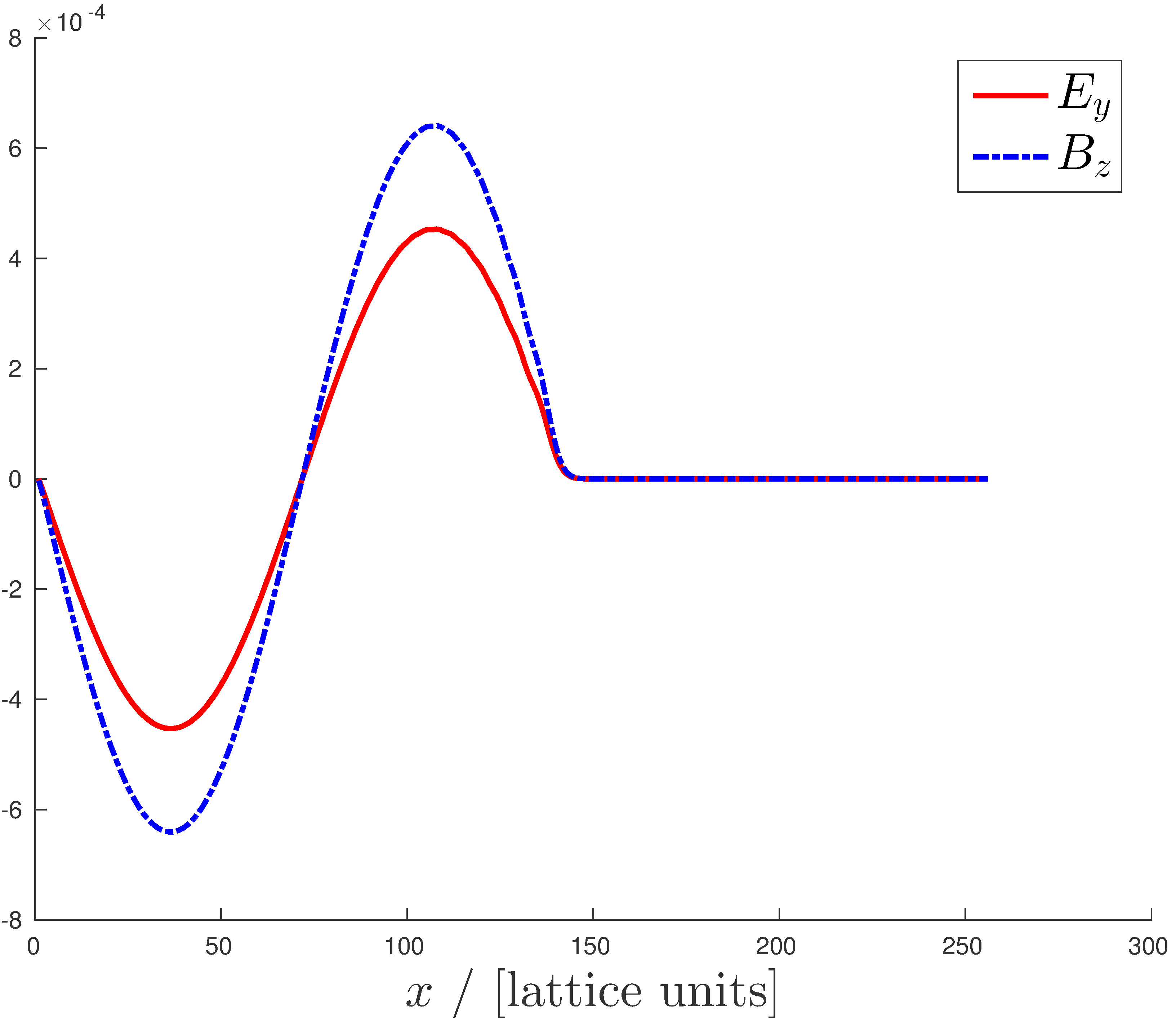}
		\caption{\textbf{Propagation of an electromagnetic wave}. Fields generated by imposing an oscillating electric field in $y-$direction at $x=0$ after a period of oscillation. The $z-$component of the magnetic field and the $y-$component of the electric field are depicted. Note that the wavefront propagates at a velocity of $1/\sqrt{2}$ (in lattice units).}
	\label{fig:speed_of_light}
\end{figure}

We chose a grid with $\delta x=\delta/\sqrt{2} t =1$, that is, the speed of electromagnetic radiation has to recover the value $c=1/\sqrt{2}$, since $\mu_0=\epsilon_0=1$. To test this on a $500\times50\times1$ grid, an oscillating electrical field $\vec{E} = \kl{0, E_y\kl{t}, 0}$ has been imposed at $x=0$, where 
\begin{equation}
 E_y\kl{t} = A_0 \sin\ekl{\omega\, t} \quad .
\end{equation}
From the simulation, see Fig. \ref{fig:speed_of_light}, one can see that over time and space the amplitude does not decay, i.e., energy is conserved. Another important feature, the relation between the fields 
\begin{equation}
 \vec{B} = \frac{n}{c} \hat{k}\times\vec{E} =\sqrt{2} \hat{k}\times\vec{E} \quad , 
\end{equation}
is recovered for the amplitude and the direction of propagation. The wavefront propagates at a velocity of $c=1/\sqrt{2}$, which is the speed of light in lattice units, as seen in Fig.~\ref{fig:speed_of_light}.

\subsection{Hall effect in classical conductors}
\label{sec:simple_hall}

For this test, an electric field $\vec{E}=\kl{E_0,0,0}$ with $E_0 \in \mkl{10^{-9}, 5\bcdot10^{-9}, 10^{-8}}$ has been applied to a channel of size $1\times100\times1$, with periodic boundary conditions in $x$ and $z$ direction. A magnetic field $\vec{B} = \kl{0,0,B_0}$ with $B_0=10^{-9}$ is imposed on the whole system as well. The electric charges have been chosen as $q_1 = -q_0 = q = 1$ and the relaxation times $\tau_0 = \tau_1 = 0.51$. $\nu$ was varied over several orders of magnitude. Simulations ran until the velocity converged to a stationary solution. 
\begin{figure}
		\centering
		\includegraphics[width=0.4\textwidth]{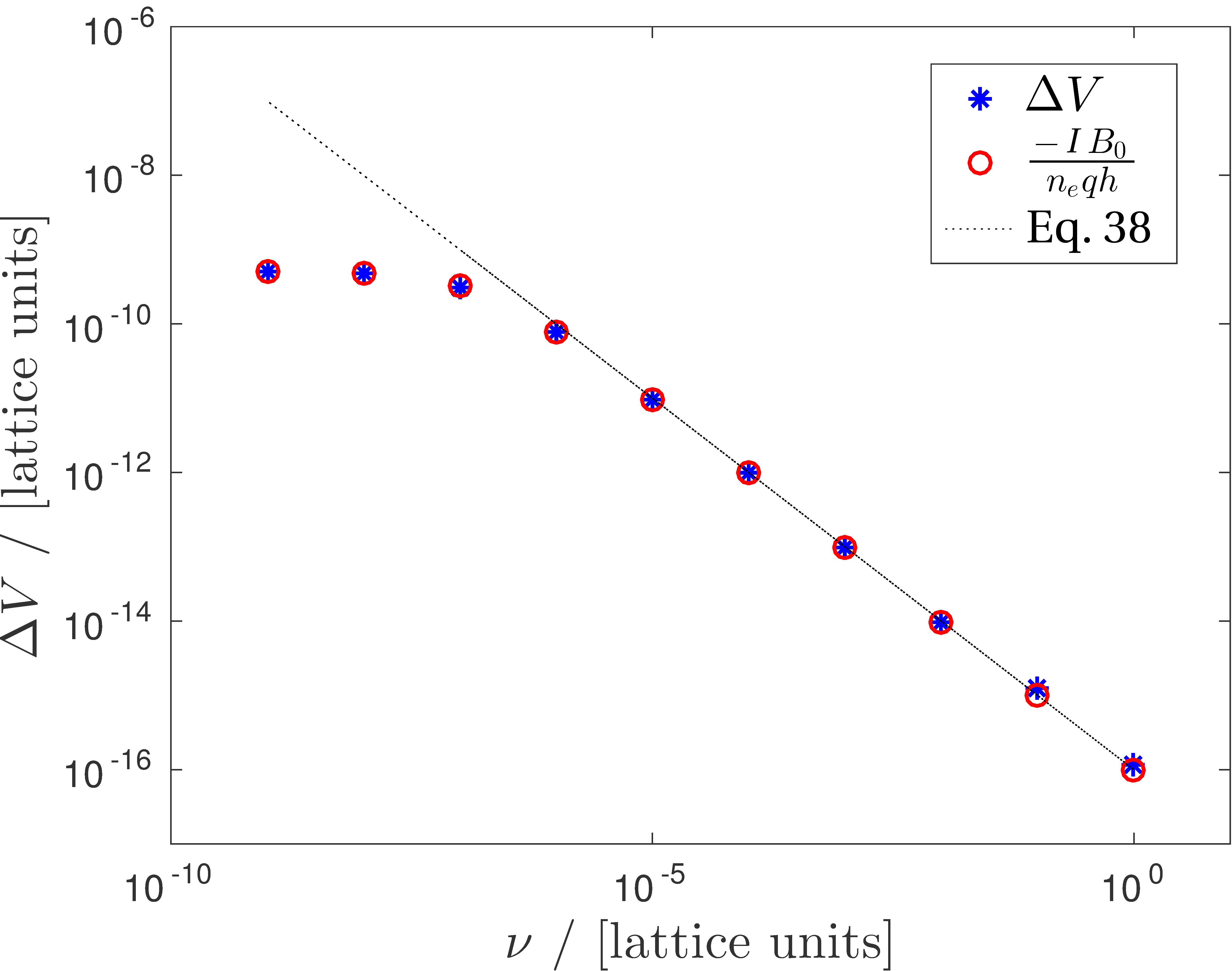}
		\caption{\textbf{Hall effect in conductors.} Hall Voltage across the conductor for different $\nu$.}
	\label{fig:hall_law_test}
\end{figure}
The expected behavior, see Eq. \eqref{eq:expl_depend2}, was recovered as it can bee seen in Fig. \ref{fig:hall_law_test}. For lower $\nu$, the system leaves the proportionality to $1/\nu$ (i.e, Eq. \eqref{eq:expl_depend}). This is because the velocity becomes so high that the viscosity term in Eq. \eqref{eq:two_fluid_navier} cannot be neglected anymore. At this point, the maximal speed is bounded by the channel flow and therefore the plateau. For the experiments of plasma flow with a transverse magnetic field we will therefore consider the current $I$ instead. This is in agreement with the expected behaviour for all the tested order of magnitudes. Doubling the system size or changing the direction of the flow or of the magnetic field does not affect the test. 

\section{Hall effect in plasmas}
\label{subsec:Hall_in_plasma}

As in classical conductors, if a constant electric field $\vec{E} = \kl{E_0,0,0}$ is applied in a Plasma with a conductivity $\sigma_0$, it will induce a current density $\vec{j} = \sigma_0 \vec{E}$. The acceleration of the charges due to a transverse field $\vec{B} = \kl{0,0,B_0}$ is proportional to the Lorentz force:
\begin{equation}
	\vec{a}_s = \frac{\vec{F}_{L, s}}{m_s}=\frac{q_s}{m_s}\kl{\vec{v}_s\times\vec{B}} \quad ,
\end{equation}
Furthermore, with the electron density $n_0$, the ion density $n_1$ and the cross section of the flow $A = h\bcdot d$, the total current in $x-$direction can be calculated as
\begin{equation}
\label{eq:current_in_hall}
	I = n_0 \,q_0 A v_{0, x}  +   n_1 \,q_1 A v_{1,x} \quad ,
\end{equation}
where $v_{s,x}$ denotes the $x$ component of the velocity $\vec{v}_s$. In the case of a classical conductor, only the electrons contribute to the current. They will drift towards one side of the conductor until a measurable voltage that balances the Lorentz force is built across the plate. Using  Eq. \eqref{eq:current_in_hall}, and then taking the limit of $m_1>>m_0$ in the generalized Ohm's law, Eq. \eqref{eq:general_ohm}, together with \eqref{eq:sigma_0}, the Hall voltage can be calculated in the following way:
\begin{align}
	\label{eq:expl_depend}
		\Delta V 	&= - \frac{I B_0}{n_0 q_0\, h} \\
	\label{eq:expl_depend2}
					&= - \frac{q_0 E_0  B_0}{ h\, m_0 \nu} \quad . 
\end{align}

In Eq. \eqref{eq:expl_depend2}, the special case of the generalized Ohm law described in Eq. \eqref{eq:simpl_ohm} has been used. From Eq. \eqref{eq:expl_depend}, we can define the dimensionless ratio
\begin{equation}
	 \mathcal{R}_{\text{H}} \equiv -\frac{\Delta V n_0 q_0 h}{I B_0}.
	\label{eq:ratio}
\end{equation}
By definition, in classical conductors $\mathcal{R}_{\text{H}} = 1$. The same ratio is expected deviate from unity if the assumptions are modified as mentioned for the case of a plasma. In particular, in this work we study the deviations of $\mathcal{R}_{\text{H}}$ in dependence of the mass ratio $\frac{m_1}{m_0}$  and the viscosity ratio $\frac{\eta_1}{\eta_0}$ of two populations of charge carriers. 

The current follows Eq. \eqref{eq:general_ohm} and as long as an electric field is applied to the plasma, the denominator in Eq. \eqref{eq:ratio} will always be bigger than zero. $\Delta V$, on the contrary, goes to zero for $\frac{m_1}{m_0}=\frac{\eta_1}{\eta_0}=1$. The effect of an increasing $\eta_1$ can be thought as slowing down the ions. Thus even if the masses are the same, for a high ionic viscosity, the electrons will move faster and a voltage can be built across the flow. For very high $\frac{m_1}{m_0}$, the movement of the ions is negligible in the contribution of the current and unity has to be recovered. 

\subsection{\label{sec:simulations}Numerical simulations}

By changing the ratio of the masses ($m_R\equiv\frac{m_1}{m_0}$) and of the viscosities ($\eta_R\equiv\frac{\eta_1}{\eta_0}$)  via tuning of $m_1$ and $\tau_1$, the effects described in Sec. \ref{subsec:Hall_in_plasma} could be confirmed. Since the flow in $y$-direction has been kept in a non-turbulent regime throughout all the simulations, the system can be restricted to a $1\times128\times1$ grid with periodic boundary conditions along the first and third dimension ($x$ and $z$). Comparisons with simulations run in a grid of $128\times128\times1$ showed no sensible difference. A constant electric field $\vec{E}_0 = \kl{E_0,0,0}$ and a transverse magnetic field $\vec{B}_0 = \kl{0,0,B_0}$ have been imposed. The simulations ran with fixed $E_0 = 10^{-9}$ and $B_0=10^{-3}$, such that the transversal velocity of the charge carriers could be neglected compared to the velocity in $x$ direction. The studied quantity is the ratio $\mathcal{R}_{\text{H}}$ described in Eq. \eqref{eq:ratio}. Convergence is reached at the iteration $i$ if the following conditions are met:
\begin{equation}
\sqrt{\frac{1}{L}\sum_{i,j,k}^{L} \kl{\frac{v_{0,x;i,j,k}\kl{i} - v_{0,x;i,j,k}\kl{i-1}}{{v}_{0,x;i,j,k}\kl{i} }  }^2 } < 10^{-8},
\label{eq:conv_chriterion_hall_exp}
\end{equation}
\begin{equation}
\abs{\frac{<\mathcal{R}_{\text{H}}\kl{i+1}>-<\mathcal{R}_{\text{H}}\kl{i}>}{<\mathcal{R}_{\text{H}}\kl{i+1}>}}    < 10^{-8} \quad ,
\label{eq:conv_chriterion_hall_exp_2}
\end{equation}
with $L \equiv L_x \times L_y \times L_z$ the system size, accordingly. This ensures that the current in $x$ direction is fully developed, since the carriers' speed is not changing anymore, and that $\mathcal{R}_{\text{H}}$ reaches a constant average value. The first $10^4$ iterations run whether the conditions are already met or not.
\begin{figure}
		\centering
		\includegraphics[width=0.48\textwidth]{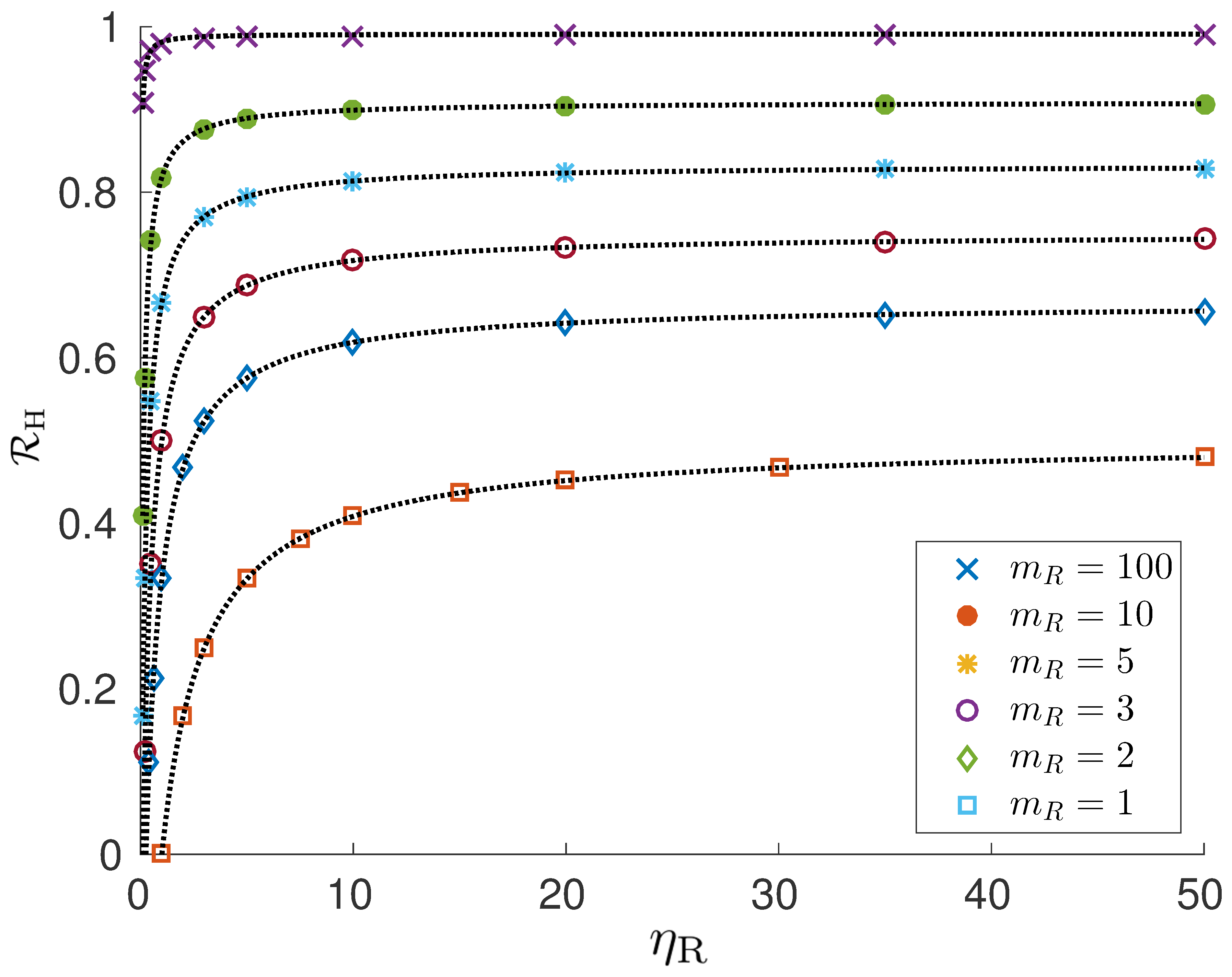}
		\caption{\textbf{Hall voltage in plasma.} Results of the simulations for $E_0 =  10^{-9}$, $B_0 =  10^{-3}$ and $\tau_0 = 0.51$. The different data points are for different mass ratios, sweeping $\eta_{\text{R}}$ from $0.1$ to $50$. The function described in Eq. \eqref{eq:fitfun} has been used for the fitting. Note that for $m_{\text{R}} = 1$ the ratio goes to zero before $\eta_{\text{R}}$ reaches zero. In fact, the simulations for equal masses yield $\Delta V = 0$ for $\eta_{\text{R}}=1$.}
	\label{fig:hall_law}
\end{figure}
Simulations have been carried out with $\tau_0=0.51$, sweeping $m_{\text{R}}$ from $1$ to $100$ and $\eta_{\text{R}}$ from $0.1$ to $50$ via changing of $\tau_1$ . Throughout all the simulations $n_0 = n_1 =1$ and  $q = 1$. In Fig. \ref{fig:hall_law}, the behavior of the ratio $\mathcal{R}_{\text{H}}$ has been depicted. It can be seen that the ratio recovers the expected behavior for large $m_R$,  for large $\eta_{\text{R}}=1$ and  for $m_{\text{R}}=\nu_{\text{R}}=1$.

For the fit, a function with asymptotic behavior has been chosen. Our ansatz is of the form
\begin{equation}\label{eq:fitfun}
\mathcal{R}_{H}  = a\kl{1 - \frac{1+ b}{1+ c \, \eta_{\text{R}} }} \quad ,
\end{equation}
where $a$ is the asymptotic value for high $\eta_{\text{R}}$ and $c$ a scaling factor for the viscosity ratio and it takes into account the fact that the ratio for equal masses goes to zero exactly when $\eta_1 = \eta_0$, and at different values for unequal masses. Additionally, $b$ represents some fitting parameter.

\begin{figure}
			\centering
			\includegraphics[width=0.48\textwidth]{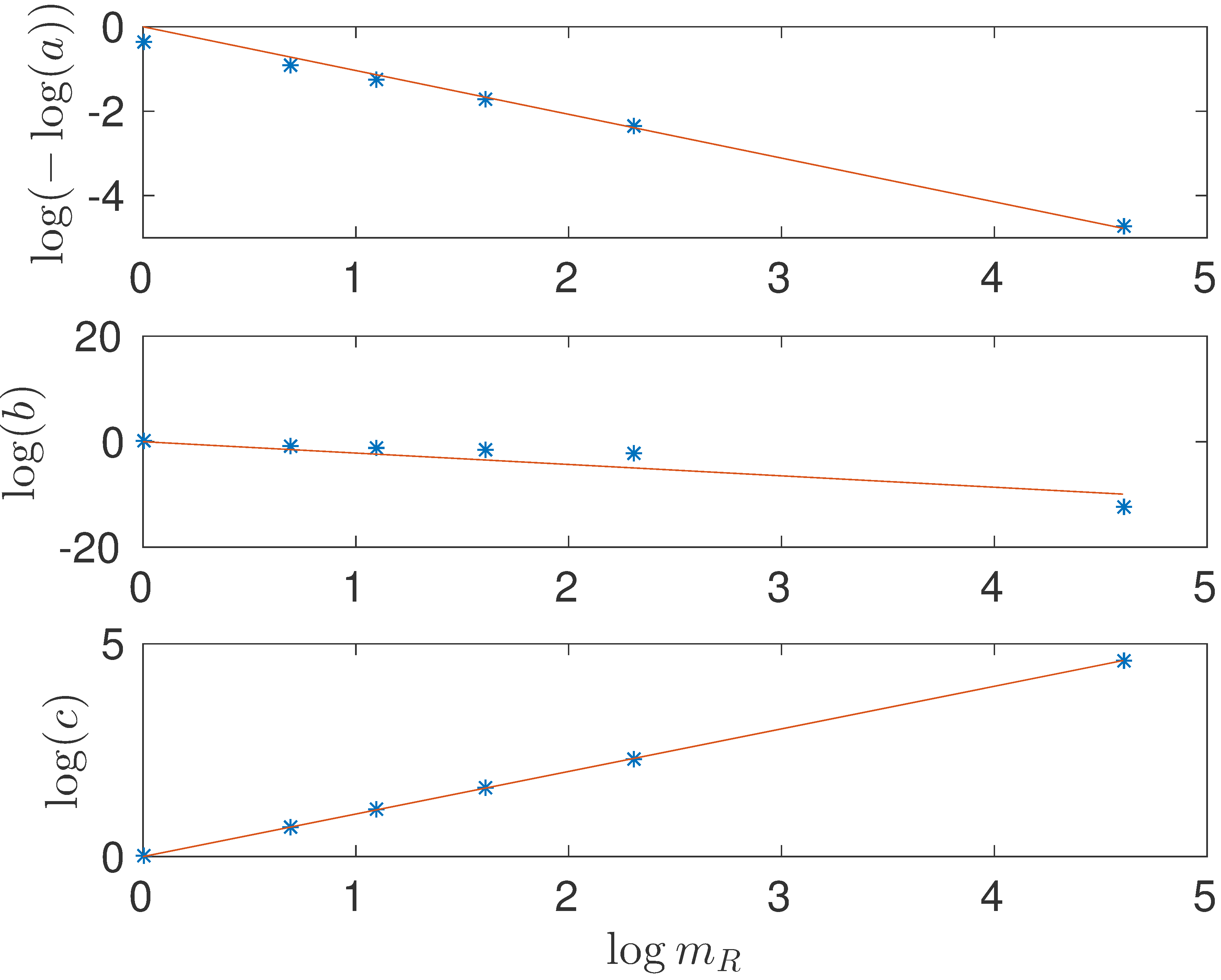}
			\caption{\textbf{Parameter fit.} Fitting of the parameters $a$, $b$ and $c$ versus $m_R$.}
			\label{fig:rParams}
\end{figure}

Comparing the parameters to the dimensionless variable $m_{\text{R}}$ shows a simple relation between these. Simulations for different $\eta_0$ consistently result the following form (See Fig. \ref{fig:rParams}):
\begin{align}  
a &= \exp\kl{-m_{\text{R}}^{-1}} \quad , \nonumber  \\ 
b &= m_{\text{R}}^{-1} \quad , \nonumber \\ 
c &= m_{\text{R}} \quad  . \nonumber \\ 
\label{eq:param_depen}
\end{align}  
Thus the Hall voltage in a plasma flow composed of two species with opposite charge and different masses can be expressed as:   
\begin{equation}
		\Delta V 	= - e^{-m_R^{-1}}\kl{1-\frac{1+m_{\text{R}} ^{-1}}{1+  m_{\text{R}} \eta_{\text{R}}  }}\frac{I B_0}{n_0 q\, h}  \quad .
	\label{eq:corrected_Hall}
\end{equation}
Note that for $m_R>>1$ and $\eta_R>>1$, Eq. \eqref{eq:corrected_Hall} recovers Eq. \eqref{eq:expl_depend}, while when $m_R\xrightarrow[]{}1$ and $\eta_{\text{R}}\xrightarrow[]{}1$, $\Delta V$ goes to zero, as expected.

\begin{figure}
		\centering
		\includegraphics[width=0.5\textwidth]{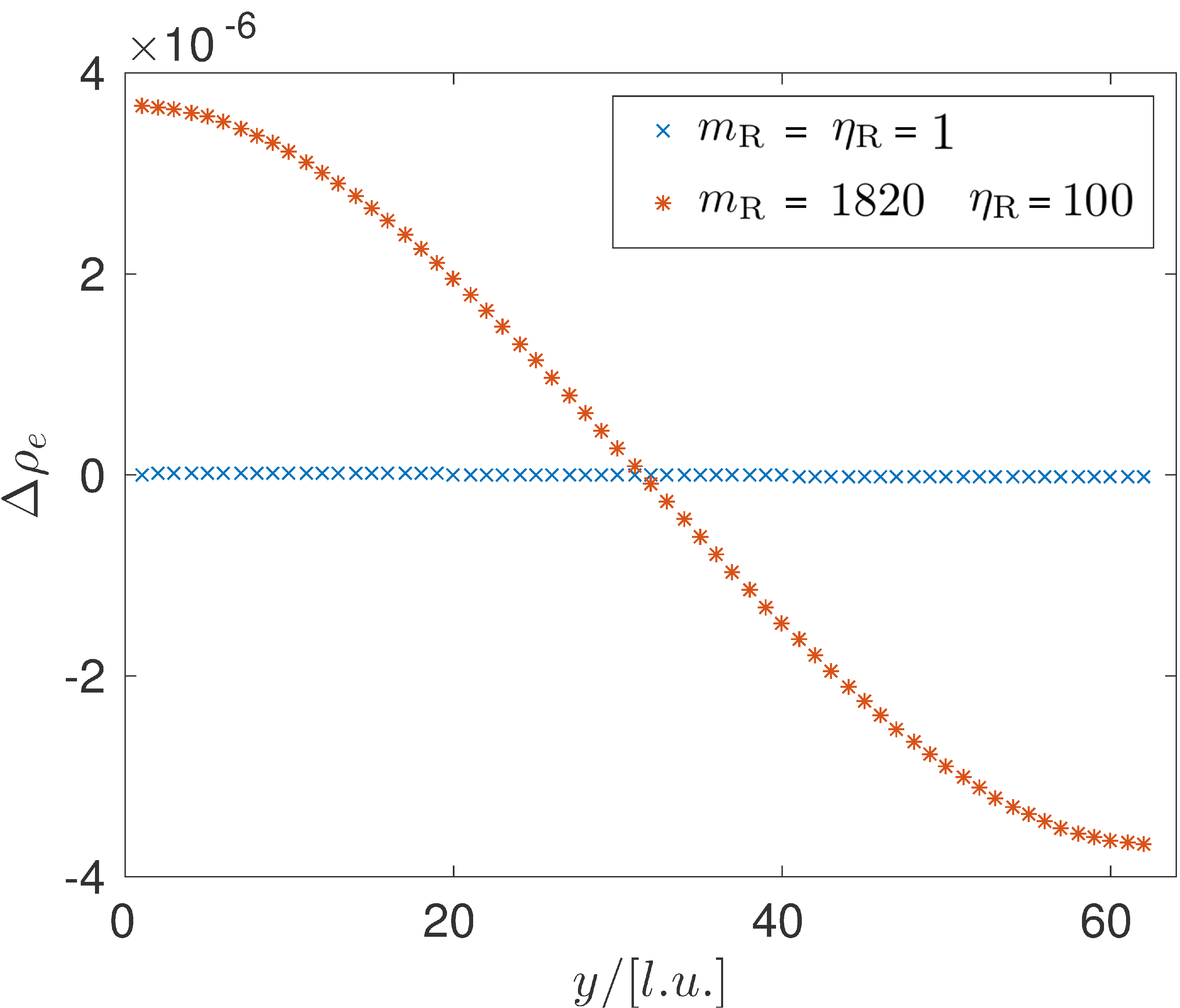}
		\caption{\textbf{Electronic density.} The density excess/defect of the electrons with respect to the average density, $\Delta \rho_0\kl{\vec{x}} \equiv \bar{\rho}_0 - \rho_0\kl{\vec{x}}$ in a $1\times64\times1$ system. For high ratios $m_{\text{R}}$ and $\eta_{\text{R}}$ the density is almost constant, while the density gradient is noticeable for low $m_{\text{R}}$ and $\eta_{\text{R}}$.}
	\label{fig:density_diff}
\end{figure}

Another interesting aspect of the Hall field in conductors is that it counterbalances the cumulation of the electrons on one side of the conductor. Since the Hall field is suppressed for  low  $m_{\text{R}}$ and $\eta_{\text{R}}$, the electrons are able to converge on one side of the channel thereby generating an appreciable gradient $\Delta \rho_0\kl{\vec{x}} \equiv \bar{\rho}_0 - \rho_0\kl{\vec{x}}$ in their density (see Fig. \ref{fig:density_diff}), with $\bar{\rho}_0$ the mean value of $\rho_0$.

\section{\label{sec:concl}Conclusions and Outlook}

We have used a lattice Boltzmann model to study the Hall effect in two-fluid plasmas. An expression, Eq. \eqref{eq:corrected_Hall}, for the cross-flow voltage in a two-fluid regime could be established. For low mass and viscosity ratios, the internal field is completely suppressed. On the contrary, for a higher mass or viscosity ratio, the voltage converges towards the value which is built up in a classical conductor. 

A main difference to the classical hall drift is that as the mass ratio and the viscosity ratio decrease, the voltage is suppressed, and the cumulation of the particles cannot  be a source of a transversal electric field since the charge density cancel itself out. The electronic and ionic pressure gradients are therefore higher than in classical conductors and a density gradient characterizes the flow.

An example for a scenario in which this effect might be of interest is in plasmas with multiple ionic charge carriers, as opposed to ionized hydrogen. Another interesting case is in the diffusion region of a magnetic reconnection event. On the separatrix lines during reconnection a differential flow creates a transversasl magnetic Hall field. This differential motion of electrons and ions is thought to trigger reconnection \cite{Zweibel20160479}. Also, non homogeneous pressure gradients in the electron population density have been shown to be not negligible for many applications in magnetic reconnection \cite{plasma_general_2,el_pres_grad2,el_pres_grad3} or turbulences induced by non-homogeneous shear in tokamaks \cite{el_pres_grad4}. Although in a very academic setting, we quantitatively established how these effects are accentuated for small mass ratios and comparable viscosities of the charge carriers.

Further interesting systems to be studied include systems in a turbulent state, in which an overall average current drives the Hall effect, and how this, in return, affects the turbulent motion of the particle species.

\begin{acknowledgments}
Financial support from the European Research Council (ERC) Advanced Grant 319968-FlowCCS is kindly acknowledged. The authors would also like to thank Ms Ashton for her useful comments to the manuscript.
\end{acknowledgments}

\bibliography{literature}



\end{document}